\documentclass[final,5p,times,twocolumn]{elsarticle}
\usepackage{hyperref}
\usepackage{epstopdf}
\usepackage{graphicx}
\usepackage{csquotes}
\begin{document}

\begin{frontmatter}

\title{Novel MPGD based Detectors of Single Photons in COMPASS RICH-1}

\author[infn-trieste]{J.Agarwala \fnref{calcutta} }
\author[torino]{M.Alexeev}                                
\author[aveiro]{C.D.R.Azevedo}
\author[infn-trieste]{R.Birsa}                                   
\author[trieste]{F.Bradamante}                                
\author[trieste]{A.Bressan}                           
\author[freiburg]{M.B\"uchele}
\author[trieste]{C.Chatterjee }
\author[torino]{M.Chiosso}                                    
\author[trieste]{P.Ciliberti}
\author[infn-trieste]{S.Dalla Torre}                   
\author[infn-trieste]{S.Dasgupta\corref{1}}
\author[infn-torino]{O.Denisov}                               
%\author[trieste]{V.Duic}                                               
\author[prague]{M.Finger}                                 
\author[prague]{M.Finger Jr.}                                         
\author[freiburg]{H.Fischer}  
%\author[trieste]{M.Giorgi}                        
\author[infn-trieste]{B.Gobbo}                      
\author[infn-trieste]{M.Gregori}                           
\author[infn-trieste]{G.Hamar}
\author[freiburg]{F.Herrmann}
%\author[freiburg]{S.Schopferer}
%\author[liberec]{D.Kramer}
%\author[freiburg]{L.Lauser}
\author[trieste]{S.Levorato}
\author[infn-torino]{A.Maggiora}
\author[trieste]{N.Makke}
\author[trieste]{A.Martin}
\author[infn-trieste]{G.Menon}
\author[prague]{J.Novy}                                 
\author[alessandria]{D.Panzieri}
\author[aveiro]{F.A.B.Pereira}
\author[aveiro]{C.A.Santos}
\author[trieste]{G.Sbrizzai}                                       
%\author[trieste]{P.Schiavon}
%\author[freiburg]{C.Schill}
\author[freiburg]{S.Schopferer}
\author[prague]{M.Slunecka}
%\author[infn-trieste]{F.Sozzi}
\author[liberec]{K.Steiger}
\author[liberec]{L.Steiger}
\author[liberec]{M.Sulc}
%\author[trieste]{S.Takekawa}
\author[infn-trieste]{F.Tessarotto}
\author[aveiro]{J.F.C.A.Veloso}
%\author[freiburg]{H.Wollny}
\author[infn-trieste]{Y.X.Zhao}

\cortext[1]{
	%Padriciano 99, 34012 Trieste, Italia. Tel. +390403756228, fax  +390403576258.
	corresponding author, email: shuddha.dasgupta@ts.infn.it}
%\fntext[dubna]{on leave from JINR, Dubna, Russia}
%\fntext[ICTP]{ICTP TRIL fellow}
\fntext[calcutta]{on leave from Matrivani Institute of Experimental Research and Education,
	Kolkata, India}
%
%\fntext[dubna]{on leave from JINR, Dubna, Russia}
%
%\address[trieste-ICTP]{INFN, Sezione di Trieste and ICTP Trieste, Italy} 
\address[infn-trieste]{INFN, Sezione di Trieste, Trieste, Italy}
\address[torino]{INFN, Sezione di Torino and University of Torino, Torino, Italy}
\address[aveiro]{I3N - Physics Department, University of Aveiro, Aveiro, Portugal}
\address[trieste]{INFN, Sezione di Trieste and University of Trieste, Trieste, Italy}
\address[freiburg]{Universit\"at Freiburg, Physikalisches Institut, Freiburg, Germany}
\address[infn-torino]{INFN, Sezione di Torino, Torino, Italy}
\address[prague]{Charles University, Prague, Czech Republic and JINR, Dubna, Russia}
\address[liberec]{Technical University of Liberec, Liberec, Czech Republic}
\address[alessandria]{INFN, Sezione di Torino and University of East Piemonte, Alessandria, Italy}

%\address[prague-ctu]{Czech Technical University, Prague, Czech Republic}
%

\begin{abstract}
COMPASS is a fixed target experiment at CERN SPS aimed to study Hadron Structure and Spectroscopy. Hadron Identification in the momentum range between 3 and 55 GeV/c is provided by a large gaseous Ring Imaging Cherenkov Counter (RICH-1). To cope with the challenges imposed by the new physics program of COMPASS, RICH-1 have been upgraded by replacing four MWPCs based photon detectors with newly developed MPGD based photon detectors. The architecture of the novel detectors is a hybrid combination of two layers of THGEMs and a MicroMegas. The top of the first THGEM is coated with CsI acting as a reflective photo-cathode. The anode is segmented in pads capacitively coupled to the APV-25 based readout. The new hybrid detectors have been commissioned during 2016 COMPASS data taking and stably operated during 2017 run. In this paper all aspects of the novel photon detectors for COMPASS RICH-1 are discussed.   
\end{abstract}


\end{frontmatter}

%\linenumbers

%%%%%%%%%%%%%%%%%%
%%				%%
%%	Section-1	%%
%%				%%
%%%%%%%%%%%%%%%%%%

\section{Introduction}

\par
\begin{figure}[h]
	\centering
	\includegraphics[width=\linewidth]{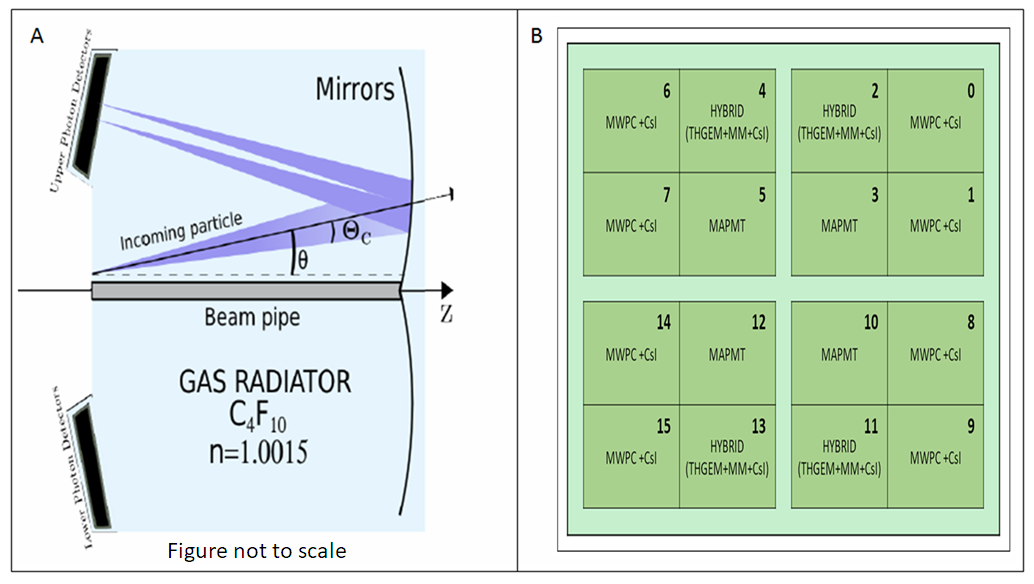}
	\caption{A. The Cherenkov photon propagation and focusing. B. Photon detectors (not to scale).}
	\label{fig:richPrinciple}
\end{figure}

 
RICH-1\cite{Albrecht} is a large gaseous Ring Image Cherenkov Counter (RICH) providing Particle Identification (PID) for hadrons within the momentum range 3 to 55 GeV/c for the COMPASS Experiment at CERN SPS\cite{compass}. It consists of a 3 m long $C_{4}F_{10}$ gaseous radiator, where charged particles with velocity above the Cherenkov threshold emit photons; 21 m$^2$ VUV mirror surface where the photons are reflected and focalized on a 5.5 $m^2$ of photo-detection surface sensitive to single photons (Fig\ref{fig:richPrinciple}-A). 
Three photo detection technologies are used in RICH-1:  Multi Wire Proportional Chambers (MWPCs) with CsI photo-cathodes, Multi Anode Photo-Multipliers Tubes (MAPMTs) and Micro Pattern Gaseous Detectors (MPGDs) based Photon Detectors (PDs) (Fig.\ref{fig:richPrinciple}-B).
%%%%%%%%%%%%%%%%%%
%%				%%
%%	Figure-1	%%
%%				%%
%%%%%%%%%%%%%%%%%%
\par
RICH-1 was designed and built in 1996-2000, commissioned in 2001-2002 and is in operation since 2002. The whole photo-detection surface was originally equipped  with 16 MWPCs with CsI photo-cathodes of $\sim 600\times600$ mm$^2$ active area. In-spite of their good performance, MWPCs have limitations in terms of maximum effective gain ($\sim 10^4$), time response($\sim \mu$s), rate capability and aging of the CsI photo-cathodes. In 2006, 4 central chambers were replaced with detectors consisting of MAPMTs coupled to individual fused silica lens telescopes to cope with the high particle rates of the central region. In parallel, an extensive R\&D program\cite{THGEM_rd} aimed to develop MPGD based large area PDs established a novel hybrid technology combining MicroMegas \cite{MM} and THick Gas Electron Multipliers (THGEMs) \cite{Alexeev}. In 2016 COMPASS RICH-1 was upgraded by replacing 4 of the remaining 12 MWPCs with CsI photo-cathodes with new detectors based on the novel MPGD hybrid technology \cite{upgradeHybrid}. The new detectors have been successfully commissioned and operated during the 2016 and 2017 COMPASS data taking periods.

%%%%%%%%%%%%%%%%%%
%%				%%
%%	Section-2	%%
%%				%%
%%%%%%%%%%%%%%%%%%

\section{The Hybrid Architecture}

%%%%%%%%%%%%%%%%%%
%%				%%
%%	Figure-2	%%
%%				%%
%%%%%%%%%%%%%%%%%%

\begin{figure}[h]
	\centering
	\includegraphics[width=\linewidth]{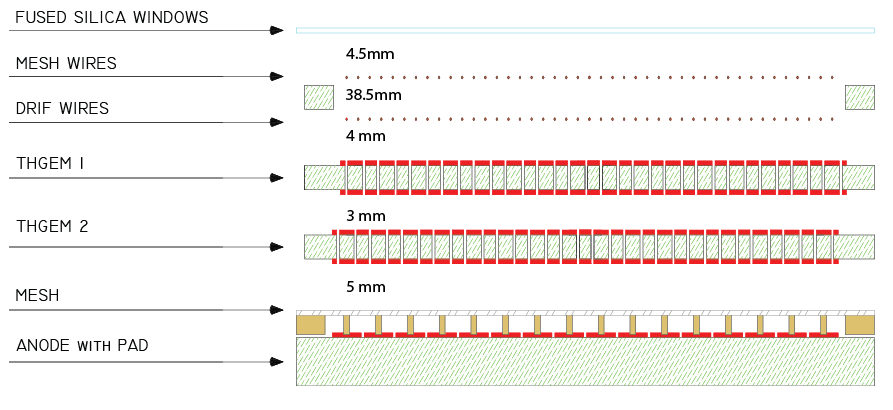}
	\caption{The Hybrid architecture}
	\label{fig:hybrid}
\end{figure}

\par
The basic structure of the hybrid module (Fig.\ref{fig:hybrid}) consists of two layers of THGEMs, one MicroMegas, and two planes of wires. UV light sensitivity is obtained via the deposit of a thin (300 nm) CsI layer on the top of the first THGEM electrode which acts as a reflective photo-cathode. 

The Drift wire plane is installed at 4 mm from the CsI coated THGEM and is biased to a suitable voltage in order to maximize the extraction and collection efficiency of the converted photo-electrons. The other wire plane guarantees the correct closure of the drift field lines and is positioned 4.5 mm away from the quartz window which separates the radiator gas volume from the $Ar:CH_{4}$ $50:50$ gas mixture of the photon detector. 

The photo-electron generated by the conversion of Cherenkov photon from the CsI surface is guided into one of the first THGEM holes where the avalanche process takes place due to the electric field generated by the biasing voltage applied between the top and bottom THGEM electrodes. The electron cloud generated in the first multiplication stage is then driven by the 1.5 $kV.cm^{-1}$ Electric field across the 3 mm transfer region to the second THGEM, where thanks to the complete misalignment of the holes with respect to the first THGEM layer ($\sim$ 462 $\mu$m displacement along the THGEM length coordinate), the charge is spread and undergoes a second multiplication process. 

%%%%%%%%%%%%%%%%%%
%%				%%
%%	Figure-3	%%
%%				%%
%%%%%%%%%%%%%%%%%%
\begin{figure}[h]
	\centering
	\includegraphics[width=0.9\linewidth]{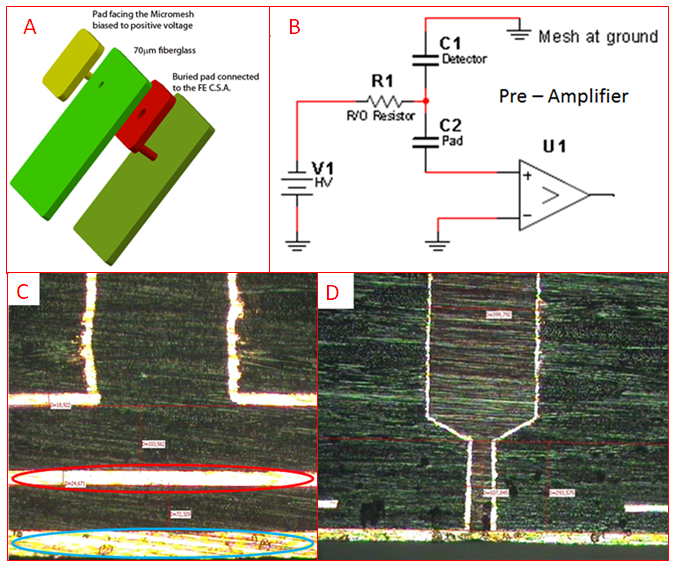}
	\caption{A: Exploded view of one single readout pad structure. B: The schematic of the circuit diagram of the \enquote{capacitive anode} idea. C and D: Metallographic section of the PCB: the detail of the through-via contacting the external pad via the hole of the buried pad. \cite{upgradeHybrid}}
	\label{fig:padscheme}
\end{figure}

Finally the charge is guided by the 0.8 $kV/cm$ field across the 5 mm gap to the bulk MicroMegas where the last multiplication occurs. The MicroMegas mesh which is the only non-segmented electrode is kept at ground potential while the anode, segmented in square pads of 7.5$\times$7.5 $mm^{2}$ (with 0.5 mm inter-pad gaps) is biased at positive voltage (Fig.\ref{fig:padscheme}-A and Fig.\ref{fig:padscheme}-B). The MicroMegas PCBs are based on the capacitive/resistive concept: the anodic pads are powered through individual resistors and the signal induced on the anodic pads is read out by the Front End APV-25 chips\cite{APV25} via capacitively coupled buried pads embedded 70 $\mu$m below the anodic ones ( Fig.\ref{fig:padscheme}-C). The high voltage is provided to the anodic pads by vias passing through the readout pads. Special attention was paid on obtaining a very flat surface for the anodic pad via connections (as shown in Fig.\ref{fig:padscheme}-D). The intrinsic ion blocking capabilities of the MicroMegas as well as the arrangements of the THGEM geometry and fields grant an ion back flow on the photo-cathode surface lower or equal to 3\% \cite{PDreview}.

%%%%%%%%%%%%%%%%%%
%%				%%
%%	Section-3	%%
%%				%%
%%%%%%%%%%%%%%%%%%

\section{Building and commissioning of the final detectors}

%%%%%%%%%%%%%%%%%%
%%				%%
%%	Figure-4	%%
%%				%%
%%%%%%%%%%%%%%%%%%

\begin{figure}[h]
	\centering
	\includegraphics[width=\linewidth]{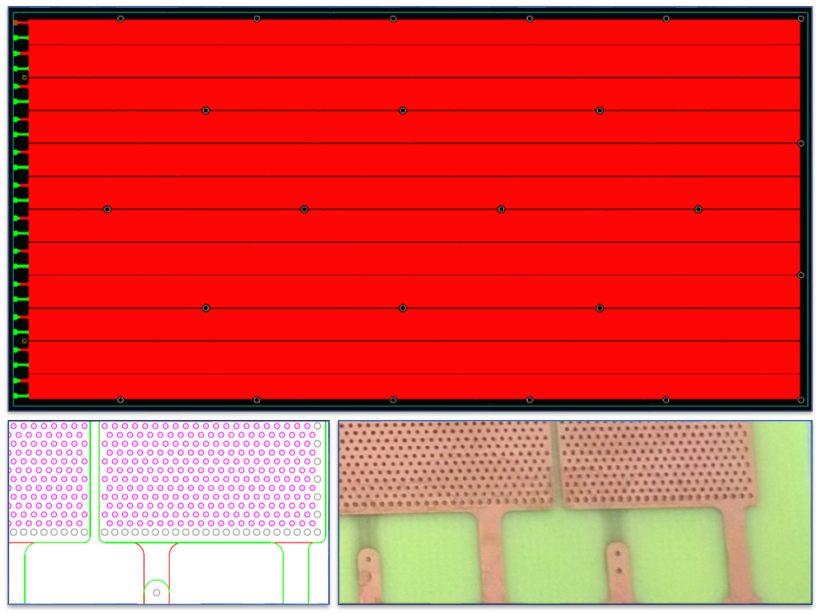}
	\caption{The final THGEM design}
	\label{fig:thgemdrawing}
\end{figure}

\par
All the THGEMs have the same geometrical parameters: thickness of 470 $\mu$m, total length of 581 mm and width of 287 mm (Fig.\ref{fig:thgemdrawing}). The holes diameter is 400 $\mu$m and the pitch is 800 $\mu$m. The holes were produced by mechanical drilling and have no rim. To obtain a symmetric field line configuration near the edges of the THGEM, the diameter of the holes located along the external borders have been enlarged to 500 $\mu$m, which results in an improved electrical stability of the whole system. The top and bottom electrodes of each THGEM are segmented in 12 sectors (10 are  564 mm long and 23.3 mm wide, the most external ones are 17.9 mm wide), all sectors are separated by 0.7 mm clearance area. The biasing voltage is individually provided to each sector of each THGEM.

\par
The THGEMs for the RICH-1 upgrade were produced following specific procedures (result of 8 year long dedicated R\&D) concerning raw material selection, THGEM production, quality assessment, characterization, storage and installation. 

\par
To achieve an effective gain uniformity of $\sim 7\%$ over large surface area ($\sim 0.2$ $m^{2}$) is challenging due to the thickness tolerance of the raw PCB sheets available in the market. To avoid waisting produced THGEMs, the selection was performed on the raw material before drilling using a setup based on MITUTOYO EURO CA776 coordinate measuring machine, at the INFN Trieste mechanical workshop. In total 50 foils were measured from which 100 THGEMs (2 THGEMs/foils) could be produced. Foils with a thickness tolerance of $\pm15$ $\mu$m ptp were sent for transfer of the mask image, etching and drilling of the holes and the others procedures needed to prepare raw THGEMs. Within 100 measured PCB pieces 60 passed the threshold and became raw THGEMs.  

\par
A post production treatment was then applied in Trieste: it consists in polishing the raw THGEM with pumice powder and cleaning by high pressurized water and in ultrasonic bath with high pH solution ($pH \sim 11$), rinsing with demineralized water and drying in oven at 50 $^{0}$C for 24 hours \cite{Polishing}. A measurement of the discharge rate was performed for the treated THGEMs using an automated test setup: in an $Ar:CO_{2}$ $70:30$ gas mixture the bias voltage of the THGEM was increased by 10 V steps and the number of sparks (events with more than 50 nA current) was measured for 30 minutes until the bias voltage was increased again. The THGEMs with the lowest discharge rates were chosen for characterization. Effective gain uniformity study was performed using a dedicated test setup consisting of a MINI-X X-Ray generator and APV-25 based SRS DAQ\cite{APV25}\cite{SRS}. The best pieces were selected for the upgrade. 

\par
The MicroMegas were produced by BULK technology at CERN EP/DT/EF/MPT workshop over the pad segmented multilayer PCBs produced by TvR SrL SpA in Schio, Vicenza, Italy. The $600\times600$ $mm^{2}$ PDs were built by mounting two $300\times600$ $mm^{2}$ modules side by side in the same aluminum frames coupled to single wire frames holding drift and field wires. The gluings of the MicroMegas PCBs to the final frames were done with the help of a volumetric dispenser coupled to a CNC machine. The assembling of each PD was performed in a clean room and it's response was studied to validate the detector. The effects of the variation of environmental conditions (pressure and temperature) were studied and an automated high voltage correction system was implemented to stabilize the PD gain response. 

\par
A special box to transport validated THGEMs under controlled atmosphere was used before and after their Au-Ni coating. The deposition of the solid photo-converter for the hybrid photo-cathodes was performed at the CERN Thin film Laboratory following the procedure described in ref-\cite{CsI}. The photo-cathodes (THGEMs with CsI coating on one side) were mounted inside a dedicated glove-box. The old PDs with MAPMT and MWPCs with CSI photo-cathodes were dismounted from RICH-1 vessel, the MAPMTs with their individual fused silica lenses were taken out from the old frames and mounted onto the frame of the new hybrid detector. The PDs are then installed on COMPASS RICH-1 and equipped with frontend electronics, low voltages, high voltages and cooling services.

%An intense pre-production and post-production quality control had been followed in the INFN THGEM Trieste Lab for all the detector pieces to build four final chambers for the upgrade. 

%%%%%%%%%%%%%%%%%%
%%				%%
%%	Section-4	%%
%%				%%
%%%%%%%%%%%%%%%%%%

\section{Results and Conclusion}

%%%%%%%%%%%%%%%%%%
%%				%%
%%	Figure-5	%%
%%				%%
%%%%%%%%%%%%%%%%%%
\begin{figure}
	\centering
	\includegraphics[width=0.7\linewidth]{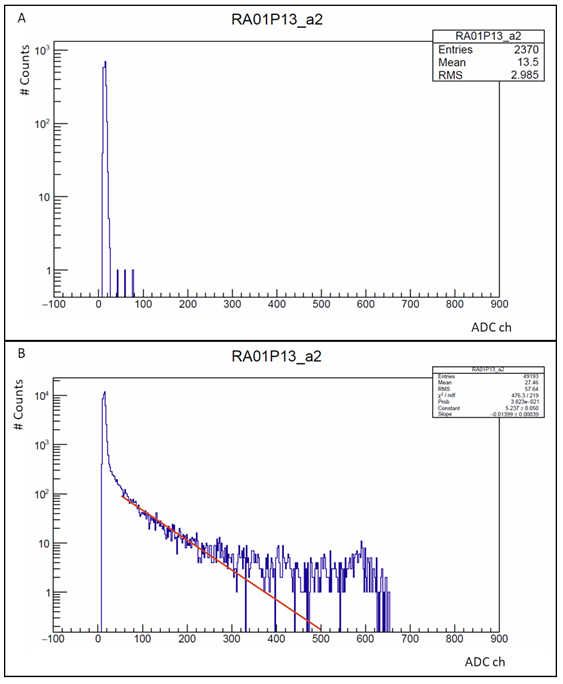}
	\caption{A. amplitude spectrum from hybrid PD13 taken without beam (random trigger); B. aplitude spectrum from hybrid PD13 taken with beam (physics trigger).}
	\label{fig:spectrawithwithoutbeam}
\end{figure}

\par
The new hybrid detectors were commissioned during the 2016 COMPASS data taking period from May to October. The average equivalent electronic noise is $\sim 900 e^{-}$ and a zero suppression procedure with a $3\sigma$ threshold cut is applied for the standard data taking. After ensuring accurate timing the amplitude spectra for noise and signals were obtained. With no beam and random trigger, the noise part of the amplitude spectrum is observed (Fig.\ref{fig:spectrawithwithoutbeam}-A). With physics triggers the amplitude spectrum shows the noise part, a prominent single photon exponential part and a tail due to charged particle signals (Fig.\ref{fig:spectrawithwithoutbeam}-B).  

%%%%%%%%%%%%%%%%%%
%%				%%
%%	Figure-6	%%
%%				%%
%%%%%%%%%%%%%%%%%%

\begin{figure}[!htb]
	\centering
	\includegraphics[scale=.65]{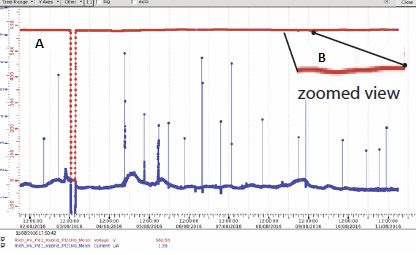}
	\caption{Voltage an current monitored for one of the hybrid bulk MicroMegas by the Hybrid HV control System for a time range of 1 week. \cite{upgradeHybrid}} 
	\label{fig:performance}
\end{figure}
The performance, in terms of voltage and gain stability of the PDs were studied. In Fig.\ref{fig:performance} the current and voltage values of a MicroMegas sector are plotted for a time range of one week. the typical discharge rate is few events per day; no sizable voltage drop is observed when a discharge occurs, confirming the validity of the detector optimization. In Fig.\ref{fig:performance} a zoomed view of the voltage curve is shown: the slow continuous voltage modulation is due to the environmental pressure temperature correction by the high voltage control system.

\par
\begin{figure}[!htb]
	\centering
	\includegraphics[scale=.40]{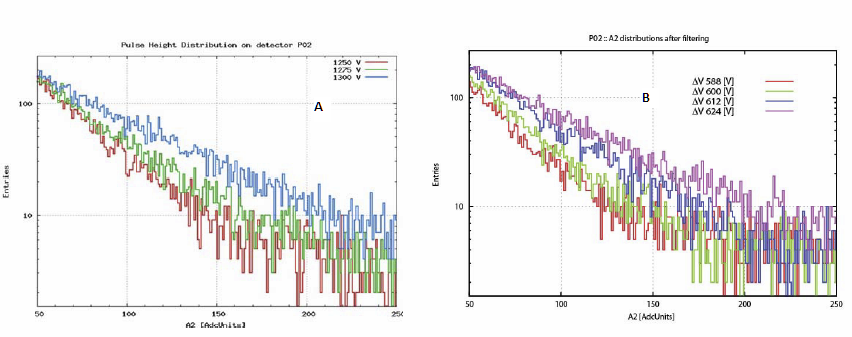}
	\caption{A. Amplitude distribution of the single photon signal collected for different biasing voltages of the second THGEM electrode: 1250, 1275,1300 V.
		B. Amplitude distribution of the single photon signal collected for different biasing voltages of the MicroMegas electrode: 588, 600, 612, 624 V. \cite{upgradeHybrid}} 
	\label{fig:spectra}
\end{figure}

The single photon amplitude spectra collected by changing only the biasing voltages of the second 
THGEM for 1250, 1275 and 1300 V are shown in figure \ref{fig:spectra} a). Similarly in fig \ref{fig:spectra} b) where only the biasing voltage of the MicroMegas is changed for 588, 600, 612, 624 V. The different slopes of the exponential distributions are in agreement with the expected values from the laboratory exercises\cite{Alexeev} and they confirm the good detector response. 
%%%%%%%%%%%%%%%%%%
%%				%%
%%	Figure-7	%%
%%				%%
%%%%%%%%%%%%%%%%%%

\begin{figure}
	\centering
	\includegraphics[width=0.7\linewidth]{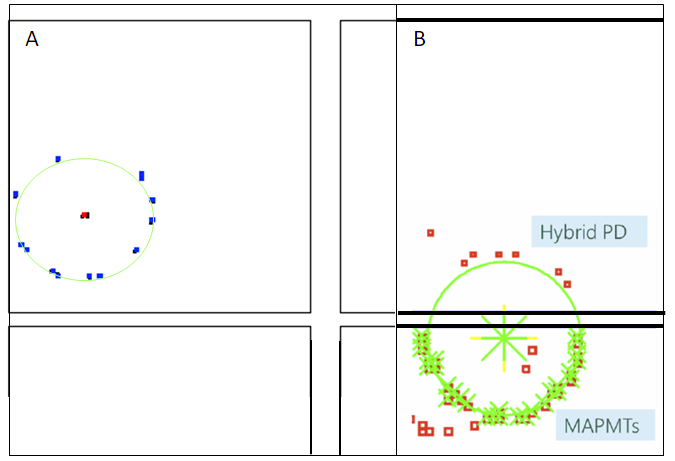}
	\caption{Cerenkov rings from RICH-1 during an typical event in COMPASS 2017 data taking}
	\label{fig:rings}
\end{figure}

Cherenkov rings are observed in the new hybrid PDs: one example is presented in Fig.\ref{fig:rings}-A where the Cherenkov photon hits are presented in blue and the red point corresponds to the extrapolated center of the expected ring, obtained by reconstructing a track (with momentum of 5.19 GeV/c in this case) and reflecting its trajectory on the mirror surface to the PDs.

\par
The four new PDs have been stably operated during 2017 COMPASS data taking periods with a higher average effective gain with respect to the MWPC based PDs. The hybrid MicroMegas + THGEMs photon-detection technology has proven to be successful and solid.
%%%%%%%%%%%%%%%%%%
%%				%%
%%	Figure-8	%%
%%				%%
%%%%%%%%%%%%%%%%%%

%\section*{References}

%\bibliography{bibfileShuddhaNDIP2017}

\section{Acknowledgment}

\par
The activity is partially supported by the H2020 project AIDA2020 GA no. 654168. It is supported in part by CERN/FIS-PAR/0007/2017 through COMPETE, FEDER and FCT (Lisbon). One author (J.Agarwala) is supported by the ICTP UNESCO TRIL fellowship program. The authors are member of the COMPASS Collaboration and part of them are members of the RD51 Collaboration: they are grateful to both Collaborations for the effective support and the precious encouragements. 

%\section*{References}

\end{document}